\documentclass[11pt,a4paper]{article}
\usepackage{amsmath}

\usepackage[psamsfonts]{amssymb}
\usepackage{amsmath}
\usepackage{epsfig}

\author{H. Mohseni Sadjadi\footnote{mohseni@phymail.ut.ac.ir}
\\ {\small Department of physics, University of Tehran ,}
\\ {\small P.O.B. 14395-547, Tehran 14399-55961, Iran}}
\title{Particle vs. future event horizon in
interacting holographic dark energy model}

\begin{document}
\maketitle
\begin{abstract}
By choosing the future event horizon as the horizon of the flat
FLRW universe, we show that the interacting holographic dark
energy model is able to explain the phantom divide line crossing.
We show that if one takes the particle event horizon as the
horizon of the universe, besides describing $\omega=-1$ crossing,
(based on astrophysical data) he is able to determine
appropriately the ratio of dark matter to dark energy density at
transition time. In this approach, after the first transition from
quintessence to phantom, there is another transition from phantom
to quintessence phase which avoids the big rip singularity.

\end{abstract}
\section{Introduction}

Based on astrophysical data, it is believed that the expansion of
the universe is accelerating \cite{astro}.  To explain the present
inflation one may assume that 70\% of the universe is composed of
a form of energy, dubbed as dark energy \cite{dark}, that
permeates all of space and has negative pressure. Many candidates
for dark energy such as cosmological constant, $\Lambda$ (vacuum
energy density) \cite{cons}, dynamical exotic scalar fields with
negative pressure \cite{scal} and so on have been introduced in
the literature. In the cosmological constant model, the dark
energy density, $\rho_{\Lambda}$, remains constant throughout the
entire history of the universe, while matter density decreases
during the expansion. So, in this model there must be a rapid
transition from matter to dark energy dominated era. This is in
contrast with the astrophysical observations which show that dark
energy and matter densities are of the same order of magnitude in
the present epoch. This is known as the coincidence problem
\cite{coinc}, which also arises in dark energy models consisting
of non-interacting exotic fields. By considering an appropriate
interaction between dark matter and dark energy components which
converts dark energy into matter, one may able to cure this
problem \cite{inter}. In \cite{inter1}, in some detail, it was
discussed how the vacuum energy density couples to the matter
fields through matter creation pressure. The process couples
cosmic vacuum (dark) energy to matter to produce future-directed
increasingly comparable amplitudes in these fields.

Some present data seem to favor an evolving dark energy with an
equation of state parameter (EOS), $\omega_d$, less than $-1$
(phantom phase) at present epoch from $\omega_d>-1$ (quintessence
phase) in the near past \cite{cross}. So it may be interesting to
take into account the possibility that the universe exhibits
phantom like behavior in the present epoch or in the future.
$\omega=-1$ crossing is not allowed in minimally coupled dark
energy models \cite{nogo}, but in multifield models,  models with
non-minimal coupling between scalar field and gravity, and in the
framework of the scalar-tensor theory of gravity this transition
is admissible \cite{nonminimal}. A question which may be arisen is
that why $\omega_d=-1$ crossing has been occurred in the present
epoch. This can be regarded as the second cosmological coincidence
problem \cite{sec}. Crossing the phantom divide line (i.e.,
crossing $\omega=-1$) may also give rise to big rip singularity
\cite{br} in a finite future time.

Based on holographic ideas \cite{cole,li}, one can determine the
dark energy density in terms of horizon radius of the universe.
Following \cite{sus}, one can show that choosing the particle
horizon as the infrared cutoff in the holographic dark energy
model, leads to a decelerating universe. In \cite{li}, the future
event horizon was examined and it was shown that this choice leads
to the expected acceleration of the universe.

To study the acceleration of the universe and the cosmological
coincidence without considering energy exchange between dark
matter and dark energy, a different approach was proposed in
\cite{simp}, by assuming that the energy density is proportional
to the inverse of the {\it area} of particle horizon within a
closed universe. Depending on the constant of proportionality,
either the ensuing inflationary period prevents the particle
horizon from vanishing, or it may lead to a sequence of big rips
as the particle horizon repeatedly traverses the closed universe.
This model does not require the decay of matter, and has a natural
reference scale, the radius of curvature.

In this paper we consider the interacting holographic dark energy
model, i.e. we assume that the amount of dark energy is
proportional to the mass of a black hole with the same radius as
the event horizon of the universe \cite{li}, and besides,  assume
that there is an interaction between dark matter and dark energy
(the (interacting) holographic dark energy model was discussed
vastly in the literature\cite{intholo}). We show that by
appropriately choosing the parameters of interaction between dark
energy and cold dark matter, one can explain $\omega=-1$ crossing
in flat Friedmann- Lemaitre- Robertson- Walker (FLRW) universe. In
this paper, $\omega$ denotes the EOS parameter of the universe. In
the first part, the future event horizon is taken as the horizon.
We show that this choice, although can describe the phantom divide
line crossing of the universe, but is inconsistent with the
thermodynamics second law for the horizon(thermodynamics of the
expanding universe has been the subject of several studies
\cite{therm}) . In addition, after the transition, the universe
remains in phantom phase and there is the possibility to encounter
the big rip singularity.

Subsequently, we show that if one takes the particle event horizon
as the horizon of the universe, besides describing $\omega=-1$
crossing, by choosing suitable interaction parameters (based on
astrophysical data), he is able to determine appropriately the
ratio of dark matter to dark energy density at transition time. In
this approach, after the first transition from quintessence to
phantom, there is another transition from phantom to quintessence
phase which avoids the big rip singularity.

We use units $\hbar=c=G=k_b=1$ throughout the paper.

\section{preliminaries}
We consider the flat FLRW metric and assume that the universe is
composed of two perfect fluids, cold (pressureless) dark matter
and dark energy. We consider exchange of energy between these
components, so they do not evolve independently:
\begin{eqnarray}\label{1}
\dot{\rho_m}+3H\rho_m&=&Q\nonumber \\
\dot{\rho_d}+3H(1+\omega_d)\rho_d&=&-Q.
\end{eqnarray}
$H$ is the Hubble parameter which in terms of the scale factor
$a(t)$ can be written as $H=\dot{a(t)}/a(t)$. The over dot
indicates the derivative with respect to the comoving time.
$\rho_m$ and $\rho_d$ are dark matter and dark energy densities
respectively. $\omega_d$ is the equation of state parameter of
dark energy. $Q$ is the interaction term which may be taken as
\cite{inter}
\begin{equation}\label{2}
Q=(\lambda_m\rho_m+\lambda_d\rho_d)H.
\end{equation}
$\lambda_m$ and $\lambda_d$ are two numerical constants. Whereas
dark matter and dark energy stress-energy tensors are not
conserved, the total stress-energy tensor is conserved:
\begin{equation}\label{3}
\dot{\rho}+3H(1+\omega)\rho=0.
\end{equation}
$\rho(=\rho_m+\rho_d)$ is the total fluid density of the universe,
in terms of which we have
\begin{eqnarray}\label{4}
H^2&=&\frac{8\pi}{3}\rho\nonumber \\
\dot{H}&=&-4\pi(1+\omega)\rho.
\end{eqnarray}
The EOS of dark energy component can be written as:
\begin{equation}\label{2000}
\omega_d=\frac{\omega}{\Omega_d},
\end{equation}
where $\Omega_d$ denotes the ratio of dark energy density,
$\rho_d$, to the total density $\rho$: $\Omega_d= \rho_d/\rho$.
Note that $0<\Omega_d<1$, therefore for $\omega\leq -1$, we have
always $\omega_d<-1$, i.e. if quintessence to phantom phase
transition occurs, the dark energy component must exhibit the
phantom like behavior but the inverse is not true: The dark energy
component may be phantom like: $\omega_d<-1$ while the universe
remains in quintessence phase, $-1<\omega<-1/3$.

Using Eqs. (\ref{1}), (\ref{2}) and (\ref{4}), one can verify that
the evolution equation of the ratio of densities of dark matter
and dark energy, denoted by $r=\rho_m/\rho_d$, satisfies
\begin{equation}\label{5}
\dot{r}=r(r+1)H\left(3\omega+\lambda_m+\frac{\lambda_d}{r}\right).
\end{equation}
Using
\begin{equation}\label{6}
\Omega_d:=\frac{\rho_d}{\rho}=\frac{1}{1+r},
\end{equation}
we find
\begin{equation}\label{7}
\omega=-\frac{1}{3H}\frac{\dot{\Omega_d}}{(1-\Omega_d)}-\frac{\lambda_d\Omega_d}{3(1-\Omega_d)}
-\frac{\lambda_m}{3}.
\end{equation}
One can consider $\rho_d$ as the holographic dark energy density
\cite{li}
\begin{equation}\label{8}
\rho_d=\frac{3}{8\pi}\frac{c^2}{L^2},
\end{equation}
where $c$ is a positive numerical constant and $L$ is the infrared
cutoff. A candidate for this cutoff is the future event horizon
defined by
\begin{equation}\label{9}
R_{f}=a(t)\int_t^\infty \frac{1}{a(t')}dt'.
\end{equation}
In the presence of big rip at $t=t_s$, $\infty$ in (\ref{9}) must
be replaced with $t_s$. Using
\begin{equation}\label{10}
\dot{R_f}=HR_f-1,
\end{equation}
which follows from (\ref{9}), and $HR_f=c/\sqrt{\Omega_d}$, we
arrive at
\begin{equation}\label{11}
\omega=-\frac{1}{3}-\frac{2}{3}\frac{\sqrt{\Omega_d}}{c}+\frac{1}
{3H}\frac{\dot{\Omega_d}}{\Omega_d}.
\end{equation}
To derive (\ref{11}) we have also used
$\omega=-1-2\dot{H}/(3H^2)$. Eqs.(\ref{7}) and (\ref{11}) may be
used to determine $\omega$ and $\dot{\Omega_d}$ in terms of
$\Omega_d$:
\begin{equation}\label{12}
\omega=-\frac{1+\lambda_d-\lambda_m}{3}\Omega_d-\frac{2}{3c}
\Omega_d^{\frac{3}{2}}-\frac{\lambda_m}{3},
\end{equation}
\begin{equation}\label{13}
\dot{\Omega_d}=H\Omega_d\left(3\omega+1+\frac{2}{c}\sqrt{\Omega_d}\right).
\end{equation}
Note that, by definition, $\Omega_d$ lies in $(0,1)$. If at a
specific point, $\dot{\Omega_d}=0$, (\ref{13}) implies that higher
derivatives of $\Omega_d$ must also be zero at that point (denoted
by the point of infinitely flatness). By considering that
$\Omega_d$ is as an analytic function, infinitely flatness may
only occur at $t\to \infty$.
\section{Future event horizon and $\omega=-1$ crossing }

In the quintessence phase we have: $-1<\omega<-1/3$, while in
phantom phase $\omega<-1$. At transition time, we have
$\omega=-1$. So if the transition is allowed then
$\omega(\Omega_d)+1$ has at least one positive root in $(0,1)$ and
$\omega$ is a decreasing function of time in the neighborhood of
this root. In terms of $u$ defined by $u=\sqrt{\Omega_d}$,
$\omega=-1$ becomes
\begin{equation}\label{14}
u^3+pu^2+q=0,
\end{equation}
where
\begin{equation}\label{15}
p=\frac{c}{2}(1+\lambda_d-\lambda_m),\,\,\,
q=\frac{c}{2}(\lambda_m-3).
\end{equation}
In order to have quintessence to phantom phase transition, the
cubic equation (\ref{14}) must have at least one root in $(0,1)$.
In addition, at transition time we must have $\dot{\omega}\leq 0$.
\begin{equation}\label{16}
\dot{\omega}=-\dot{\Omega_d}\left(\frac{2p}{3c}+\frac{\sqrt{\Omega_d}}{c}\right),
\end{equation}
leads to
\begin{equation}\label{17}
\dot{\omega}=-2Hu^2\left(\frac{u}{c}-1\right)\left(\frac{2p}{3c}+\frac{u}{c}\right),
\end{equation}
at $\omega=-1$. Therefore $\dot{\omega<0}$, at transition time, is
satisfied when
\begin{equation}\label{18}
0<u<Minimum \{c,-\frac{2p}{3}\},
\end{equation}
or
\begin{equation}\label{19}
Maximum\{c,-\frac{2p}{3}\}<u<1.
\end{equation}

To go further let us discuss the behavior of the future event
horizon at transition time, denoted by $t=t_0$. In the
neighborhood of $t_0$,  we have $H(t)=
H(t_0)+h_{\alpha}(t-t_0)^{\alpha}$, where
$h_{\alpha}=(1/\alpha!)H^{(\alpha)}(t_0)$, and $\alpha$ is the
order of the first nonzero derivative of $H$ at $t_0$:
$H^{(\alpha)}=d^{\alpha}H/dt^{\alpha}>0$. Note that $\alpha$ is an
even positive integer and $h_{\alpha}$ is positive \cite{moh1}. In
a universe which will remain in phantom phase, the future event
horizon is a non increasing function of time: $\dot{R_f}\leq 0$
\cite{moh2}. Using (\ref{10}) , we obtain the following expansions
for $R_f$:
\begin{equation}\label{20}
R_f(t)=R_f(t_0)\left(1+\frac{h_\alpha}{\alpha+1}(t-t_0)^{\alpha+1}\right),
\end{equation}
provided that $\dot{R_f}(t_0)=0$, and
\begin{equation}\label{21}
R_f(t)=R_f(t_0)+\left(R_f(t_0)H(t_0)-1\right)(t-t_0),
\end{equation}
when $\dot{R_f}(t_0)\neq 0$. In the case (\ref{20}), we have
$\dot{R_f}(t)=R_f(t_0)h_\alpha (t-t_0)^\alpha$, therefore for
$t>t_0$, $\dot{R_f}(t)>0$ which is conflict with the fact that in
phantom era the future event horizon is non increasing. Hence at
quintessence to phantom transition time we must have
$\dot{R_f}\neq 0$ which, by considering $\dot{R_f}(t>t_0)\leq 0$
leads to $\dot{R_f}(t=t_0)=H(t_0)R_f(t_0)-1<0$, therefore using
the continuity of $R_f$ we conclude that there exists a
neighborhood, $N$ of $t_0$, for which  $\dot{R_f}(t\in N)<0$ .

It should be noted that the validity of the above results is
contingent on the fact that the universe will remain in phantom
phase for all future times, i.e. if there is another transition
from phantom to quintessence phase, we may have $\dot{R_f}>0$,
even in phantom era. If this situation is allowed, (\ref{14}) must
have two roots in $(0,1)$. Therefore  Rolle's theorem implies that
there exists a $t\in(0,1)$ such that $\dot{\omega(t)}=0$. At this
time we have $u=-2p/3$ and
$\ddot{\omega}=-{\dot{\Omega_d}}^2/(2c\sqrt{\Omega_d})<0$ which
follows from (\ref{12}). Therefore $\omega$ is a concave function
and the (possibly) allowed successive transitions are phantom
$\to$ quintessence $\to$ phantom. Hence we conclude that even in
the presence of two transitions, after the quintessence to phantom
transition the system remains in phantom phase. On the other hand,
by considering $HR_f=c/u$, $\dot{R_f}\geq 0$ is satisfied when and
only when $u\leq c$ (see (\ref{10})), therefore following the
above discussion which implies that at transition time
$\dot{R_F}<0$, only the choice (\ref{19}) is allowed. (\ref{19})
also implies that there is a lower bound for dark energy density
at transition time.

\begin{figure}
\centering\epsfig{file=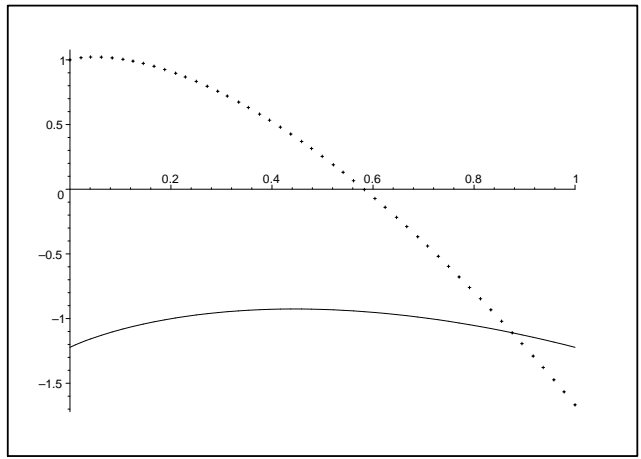,width=6cm,angle=270}
\caption{$\omega$ as a function of $\Omega_d$, for
$\lambda_m=11/3$, $\lambda_d=-10/3$ and $c=1/3$(continuous line)
and for $\lambda_m=-3$, $\lambda_d=-8$ and $c=1/6$ (points).}
\label{fig1}
\end{figure}

As an example let us assume $Maximum\{c,-2p/3\}=-2p/3$. In this
case, at transition time we must have $-2p/3<u<1$.  In order that
the transition occurs the cubic polynomial $Q(u):=u^3+pu^2+q$,
must have at least a positive root in $(-2p/3,1)$. Following
Descartes sign rule, for $p\geq 0,\,\,q<0$ and $p<0,\,\, q\leq 0$,
$Q(u)$ has a positive root, while for $p<0,\,\, q>0$ it has either
two or no positive roots.  Consider the sequence $D=\{u^3+pu^2+q,
3u^2+2pu, 6u+2p, 6\}$. We have
\begin{eqnarray}\label{23}
D(1)&=&\{1+p+q, 3+2p, 6+2p,6\}\nonumber \\
D(-\frac{2p}{3})&=&\{\frac{4p^3}{27}+q,0,-2p,6\}.
\end{eqnarray}
By Applying Budan-Fourier theorem and Descartes sign rule (and
considering $0<-2p<3$ which results in $1+p>4p^3/27$), we conclude
that $Q$ at most has one root in $(-2p/3,1)$, provided
\begin{equation}\label{24}
\frac{4p^3}{27}+q<0,\,\, 1+p+q\geq 0.
\end{equation}
As an illustration, the plot of $\omega$ is depicted for
$(p=-1/3,q=-1/2,c=1/6)$ corresponding to $(\lambda_m=-3,
\lambda_d=-8, c=1/6 )$ and $(p=-1,q=1/9,c=1/3)$ corresponding to
$(\lambda_m=11/3, \lambda_d=-10/3$ and $c=1/3)$ in
fig.(\ref{fig1}). For $(p=-1/3,q=-1/2,c=1/6)$, $q$ is negative and
$\omega=-1$ has only one root which lies in $(-2p/3,1)$. In this
case the transition is quintessence $\to$ phantom. For
$(p=-1,q=1/9,c=1/3)$, $q$ is positive and $\omega=-1$ has two
roots in $(0,1)$. One of these roots lies in $(-2p/3,1)$. Note
that in this case as we have verified previously, the transitions
are phantom $\to$ quintessence $\to$ phantom.

At the end of this section it is worth to note that if we assign
an entropy to the future event horizon via
\begin{equation}\label{22}
S=\pi R_f^2,
\end{equation}
our previous discussions reveal that the second thermodynamics law
for the horizon is not respected at least in the transition epoch:
$\dot{S}(t\simeq t_0)<0$.  Note that $\dot{S}(t>t_0) \leq 0$ in
phantom phase. The above entropy may be assumed as the entropy of
the whole system \cite{li,secondlaw}, indeed in the holographic
approach, all information stored within some volume is represented
on the boundary of that region.

In another point of view the entropy attributed to the horizon is
considered as a measure of our ignorance about what is going
behind it.  In this approach the total entropy of the universe,
$S_T$, is the sum of the entropy of perfect fluids inside the
horizon, $S_m$,  and the horizon entropy, $S_T=S+S_m$. In this way
the validity of the generalized second law (GSL) of
thermodynamics, i.e $\dot{S_T}\geq 0$, may be investigated
\cite{gsl}. Recently in \cite{moh1} it has been shown that, at
least, at transition epoch GSL is not respected.

In brief in this section we have shown that although by taking the
future event horizon as the infrared cutoff, the holographic dark
energy model can describe $\omega=-1$ crossing. Besides, in this
model, the universe will remain in phantom phase after the
transition(s) and the cosmological evolution may be ended by the
big rip singularity.

\section{Particle horizon and twice $\omega=-1$ crossing}

In this section we consider the particle horizon defined by
\begin{equation}\label{25}
R_p=a\int_0^t\frac{dt'}{a(t')},
\end{equation}
as the infrared cutoff and study $\omega=-1$ crossing in
accelerated expanding universe. In this situation, (\ref{12}) and
(\ref{13}) must be replaced with
\begin{equation}\label{26}
\omega=-\frac{1+\lambda_d-\lambda_m}{3}\Omega_d+\frac{2}{3c}
\Omega_d^{\frac{3}{2}}-\frac{\lambda_m}{3},
\end{equation}
and
\begin{equation}\label{27}
\dot{\Omega_d}=H\Omega_d\left(3\omega+1-\frac{2}{c}\sqrt{\Omega_d}\right),
\end{equation}
respectively. For $\omega<-1/3$, i.e. for accelerated expanding
universe, $\Omega_d$ is a decreasing function of comoving time. In
terms of $p$ and $q$ defined in (\ref{15}), we have
\begin{equation}\label{28}
\omega+1\equiv\frac{2}{3c}(u^3-pu^2-q).
\end{equation}
Following Descartes sign rule, the  equation $\omega+1$ may have
two positive real roots only when $p>0,\,\,q<0$. If the equation
$\omega+1=0$ has two roots in $(0,1)$, following Rolle's theorem
we expect that $\dot{\omega}(t)=0$ for a $t$ between the roots.
$\dot{\omega}=0$ occurs at $u=2p/3$, and at this point
$\ddot{\omega}=\dot{\Omega_d}^2/(2c\sqrt{\Omega_d})>0$, which
shows that $\omega(t)$ is convex function. At $\omega=-1$,
\begin{equation}\label{29}
\dot{\omega}=2Hu^2\left(-1-\frac{u}{c}\right)\left(-\frac{2p}{3c}+\frac{u}{c}\right).
\end{equation}
Transition from quintessence (phantom) to phantom (quintessence)
occurs when $\dot{\omega}<(>)0$, which following (\ref{29}) leads
to $u_0>(<)2p/3$, where $u_0$ is the root of (\ref{28}). Based on
the above discussion we conclude that in order to have two
transitions, one from quintessence to phantom and the other from
phantom to quintessence,  the equation (\ref{28}) must have two
roots. One of these roots lies in $(0,2p/3)$ and the other must be
located in $(2p/3,1)$.  The minimum of $\omega$ occurs at $u=2p/3$
in phantom era: at this point  $\dot{\omega}=0$ and
$\ddot{\omega}>0$.

Consider the sequence $D(u)=\{ u^3-pu^2-q, 3u^2-2pu, 6u-2p, 6\}$,
we have $D(0)=\{-q,0,-2p,6\}$, $D(1)=\{1-p-q, 3-2p, 6-2p, 6\}$ and
$D(2p/3)=\{-4p^3/27-q,0,2p,6\}$. The discriminant of the cubic
polynomial, $u^3-pu^2-q$, is: $-q(27q+4p^3)$. If we expect that
this polynomial has two roots in $(0,1)$, the third root must also
be real ($p$ and $q$ are real) and therefore the discriminant must
be positive. But as we have seen before, $q<0$, which results in
$-4p^3/27-q<0$. Therefore there is only one sign change in
$D(2p/3)$. In $D(0)$, for $p>0\,\,q<0$ there are two sign changes.
So if we assume $1-p-q>0$, there is no sign change in $D(1)$ (
note that $2p/3<1$ ) and following Budan-Fourier theorem we
conclude that there is one root in $(0,2p/3)$ and the other root
is in $(2p/3,1)$. In brief if $p$ and $q$ satisfy
\begin{equation}\label{30}
0<p<1.5,\,\,\, q<0,\,\,\, \frac{4p^3}{27}+q>0, \,\,\,1-p-q>0
\end{equation}
the transitions quintessence $\to$ phantom  $\to$ quintessence are
allowed. In this case besides that the big-rip is avoided, the
thermodynamics second law for the horizon is also respected:
$\dot{S}>0$, which follows from
\begin{equation}\label{31}
\dot{R_p}=HR_p+1>0.
\end{equation}
At phantom to quintessence transition time, $u$ has a lower bound:
$2p/3$, and at phantom to quintessence transition time we have
$0<u<2p/3$. In the accelerating regime, $\omega<-1/3$, and $u$ is
a decreasing function of time.

It is worth noting that there is no sign change in the sequence
$D(p)=\{-q,p^2,4p,6\}$. So the root of (\ref{28}), corresponding
to the ratio of dark energy density to total energy density at
quintessence to phantom transition, must be restricted to $u\in
(2p/3,p)$. Thus by choosing an appropriate $p<1$, one can
determine $r$, in agreement with astrophysical data.

\begin{figure}
\centering\epsfig{file=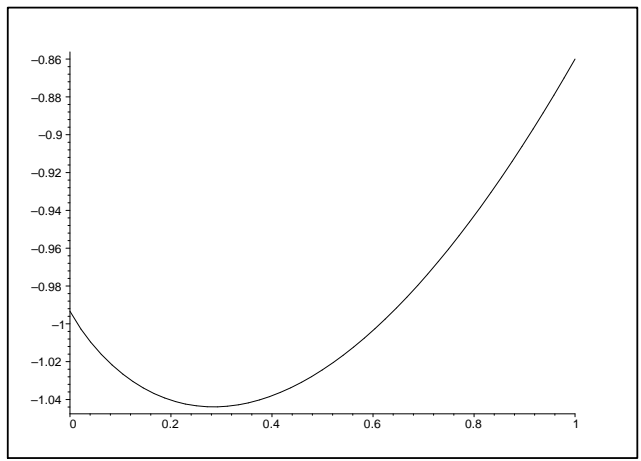,width=6cm,angle=270}
\caption{$\omega$ as a function of $\Omega_d$, for
$\lambda_m=2.98$, $\lambda_d=3.58$ and $c=1$.} \label{fig2}
\end{figure}

As an illustration, $\omega$ is depicted in fig.(\ref{fig2}) for
$p=0.8, q=-0.01$ corresponding to $\lambda_m=2.98, \lambda_d=3.58,
c=1$. The transitions occur in $u=0.78(\Omega_d=0.61)$ and
$u=0.12(\Omega_d=0.014)$ for quintessence $\to$ phantom and
phantom $\to$ quintessence respectively.

\newpage


\begin{thebibliography}{99}
\bibitem{astro} A. G. Riess et al. (Supernova Search Team Collaboration), Astron. J.
116, 1009 (1998); S. Perlmutter et al., Nature (London) 391, 51
(1998); S. Perlmutter et al. (Supernova Cosmology Project
Collaboration), Astrophys. J. 517, 565 (1999).
\bibitem{dark} D. Huterer and M. S. Turner, Phys. Rev. D 60, 081301
(1999).
\bibitem{cons} S. Weinberg,  Rev. Mod. Phys. 61, 1 (1989);
S. M. Carroll, Living Rev. Relativity 4, 1 (2001); V. Sahni and A.
A. Starobinsky, Int. J. Mod. Phys.  D 9, 373 (2000); E. J.
Copeland, M. Sami and Shinji Tsujikawa, Int. J. Mod. Phys. D 15,
1753 (2006).
\bibitem{scal} G. Huey, L. Wang, R. Dave, R. R. Caldwell and P. J.
Steinhardt,  Phys. Rev. D 59, 081301 (1999); S. M. Caroll, Phys.
Rev. Lett. 81, 3067 (1998); R. R Caldwell, Phys. Lett. B 545, 23
(2002); T. Chiba, T. Okabe and M. Yamaguchi, Phys. Rev. D 62,
023511 (2000);  K. Dimopoulos and J. W. F. Valle, Astropart. Phys.
18, 287 (2002); S. Tsujikawa, Phys. Rev. D 72, 083512 (2005); Z.
Guo, N. Ohta and Y. Zhang, astro-ph/0603109;  M. Alimohammadi and
H. Mohseni Sadjadi, Phys. Rev. D 73, 083527 (2006); W. Fang, H. Q.
Lu and Z.G.Huang, hep-th/0610188; X. Zhang, Phys. Rev. D 74,
103505 (2006); Z. G. Huang, H. Q. Lu and W. Fang, hep-th/0610018.
\bibitem{coinc}I. Zlatev, L. Wang and P. J. Steinhardt, Phys. Rev. Lett. 82, 896
(1999); N. Arkani-Hamed, L. J. Hall, Ch. Kolda and H. Murayama,
Phys. Rev. Lett. 85, 4434 (2000);  S.A. Bludman and M. Roos, Phys.
Rev. D 65, 043503 (2002); H. Mohseni Sadjadi and M. Alimohammadi,
Phys. Rev. D 74, 103007 (2006).
\bibitem{inter}W. Zimdahl, D. Pavon, L. P. Chimento, Phys. Lett. B 521, 133
(2001);  G. Mangano, G. Miele and V. Pettorino, Mod. Phys. Lett. A
18, 831 (2003); L. P. Chimento, A. S. Jakubi, D. Pavon and W.
Zimdahl, Phys. Rev. D 67, 083513 (2003); G. Olivares, F.
Atrio-Barandela and D. Pavon, Phys. Rev. D 74, 043521 (2006); Z.
Guo, R. Cai, Y. Zhang, JCAP 0505, 002 (2005).

\bibitem{inter1} M. R. Mbonye, Mod. Phys. Lett. A 19, 117 (2004).
\bibitem{cross}  U. Alam, V. Sahni, A.A. Starobinsky, JCAP
0406, 008 (2004); B. Feng, X. L. Wang, X. M. Zhang, Phys. Lett. B
607, 35 (2005); D. Huterer and A. Cooray, Phys. Rev. D 71, 023506
(2005); S. Nojiri and S. D. Odintsov, Phys. Rev. D 72, 023003
(2005); Ming-zhe Li, B. Feng, X. Zhang, JCAP 0512, 002 (2005); H.
Wei, N. Tang and S. N. Zhang, astro-ph/0612746; S. Y. Vernov,
astro-ph/0612487; H. Mohseni Sadjadi and M. Alimohammadi, Phys.
Rev. D 74, 043506 (2006); M. Alimohammadi and H. Mohseni Sadjadi,
gr-qc/0608016; H. Zhang and Z. Zhu, astro-ph/0611834; Z. Guo, Y.
Piao, X. Zhang and Y. Zhong Zhang, Phys. Rev. D74, 127304 (2006);
E. O. Kahya and V. K. Onemli, gr-qc/0612026; V. K. Onemli and R.
P. Woodard, Phys. Rev. D 70, 107301 (2004);  Z. Guo, Y. Piao, X.
Zhang, Y. Zhang, Phys. Lett. B 608, 177 (2005); Y. Cai, H. Li, Y.
Piao and X. Zhang, gr-qc/0609039.
\bibitem{nogo}A. Vikman, Phys. Rev. D 71, 023515 (2005).
\bibitem{nonminimal} H. Stefancic, Phys. Rev. D 71,
124036 (2005); R. R. Caldwell and M. Doran, Phys. Rev. D 72,
043527 (2005); Z. Guo, R. Cai and Y. Zhang, JCAP 05, 002 (2005);
S. Nojiri, S. D. Odintsov, and S. Tsujikawa, Phys. Rev. D 71,
063004 (2005); B. Gumjudpai, T. Naskar, M. Sami and S. Tsujikawa,
JCAP 06, 007 (2005); Z.-K. Guo, Y.-S. Piao, X. Zhang, and Y.-Z.
Zhang, Phys. Lett. B 608, 177 (2005); W. Hu, Phys. Rev. D 71,
047301 (2005); H. Wei, R. G. Cai and D. F. Zeng, Class. Quant.
Grav. 22, 3189 (2005); R. Gannouji, D. Polarski, A. Ranquet and A.
A. Starobinsky, JCAP 0609, 016 (2006).
\bibitem{sec} H. Wei and  R. Cai, Phys.Rev. D 73, 083002 (2006); H. Wei and R.
Cai, Phys.Lett. B 634, 9 (2006).
\bibitem{br}R. R. Caldwell, M. Kamionkowski, and N. N. Weinberg,
Phys. Rev. Lett. 91, 071301 (2003); F. Briscese, E. Elizalde, S.
Nojiri and S. D. Odintsov, hep-th/0612220.
\bibitem{cole}A. G. Cohen, D. B. Kaplan and A. E. Nelson, Phys. Rev. Lett. 82, 4971
(1999).
\bibitem{li}M. Li, Phys. Lett. B 603, 1 (2004).
\bibitem{sus}S. D. H. Hsu, Phys. Lett. B 594, 13 (2004).
\bibitem{simp}F. Simpson, astro-ph/0609755.
\bibitem{intholo} D. Pavon and  W. Zimdahl,
hep-th/0511053; D. Pavon and W. Zimdahl, Phys. Lett. B 628, 206
(2005); B. Guberina, R. Horvat and H. Nikolic, astro-ph/0611299
(to appear in JCAP); H. Mohseni Sadjadi and M. Honardoost,
gr-qc/0609076 (to appear in Phys. Lett. B); K. H. Kim, H. W. Lee
and Y. S. Myung, gr-qc/0612112; X. Zhang and F. Wu, Phys. Rev. D
72, 043524 (2005); X. Zhang, Int. J. Mod. Phys. D 14, 1597 (2005);
M. R. Setare, Phys. Lett. B 642, 1 (2006); B. Wang, J. Zang, C.
Lin, E. Abdalla and  S. Micheletti, astro-ph/0607126; W. Zimdahl
and D. Pavon, astro-ph/0606555.

\bibitem{therm} P. C. W. Davies, Class. Quant. Grav. 4, L225
(1987), ibid. 5, 1349 (1988); R. Bousso, Phys. Rev. D 71, 064024
(2005); S. Nojiri and S. D. Odinstov, Gen. Rel. Grav. 38, 1285
(2006); I. Brevik, S. Nojiri, S. D. Odintsov and L. Vanzo, Phys.
Rev. D70, 043520 (2004); P. F. Gonzalez-Diaz and C. L. Siguenza,
Nucl. Phys. B 697, 363 (2004); E. Babichev, V. Dokuchaev, and Y.
Eroshenko, Phys. Rev. Lett. 93, 021102 (2004); S. Nojiri and S. D.
Odintsov, Phys. Rev. D 70, 103522 (2004); M. R. Setare and  S.
Shafei, JCAP 09, 011 (2006); M. Akbar and  R. Cai, hep-th/0609128.


\bibitem{secondlaw}G. Huang and M. Li, JCAP 0408,
013 (2004); Y. Gong, B. Wang and Y. Zhang, Phys. Rev. D 72, 043510
(2005); B. Wang, Y. Gong and E. Abdalla, Phys. Lett. B 624, 141
(2005); B. Wang, C. Lin and E. Abdalla, Phys. Lett. B 637, 357
(2006).
\bibitem{gsl} R. Brustein, Phys. Rev. Lett. 84, 2072 (2000); P. C. W. Davies and T. M. Davis,  Foundations of Physics, 32(12),
1877 (2002);  G. Izquierdo and D. Pavon, Phys. Lett. B 633, 420
(2006);  B. Wang, Y. Gong and E. Abdalla, Phys. Rev. D 74, 083520
(2006); Y. Gong, B. Wang and A. Wang, JCAP 0701, 024 (2007); Y.
Gong, B. Wang and A. Wang, gr-qc/0611155.
\bibitem{moh1}H. Mohseni Sadjadi, Phys. Lett. B 645, 108 (2007).
\bibitem{moh2} H. Mohseni Sadjadi, Phys. Rev. D 73, 063525 (2006).












\end{thebibliography}
\end{document}